\documentclass[traditabstract]{aa} 
\usepackage{graphicx}
\usepackage{txfonts}
\usepackage{amssymb}
\usepackage{supertabular}
\usepackage{longtable}
\usepackage{rotating}
\usepackage{epstopdf}
\usepackage{lscape}

\begin{document}
\title{The low-mass diskless population of Corona Australis}


\author{
Bel\'en L\'opez Mart\'{\i}\inst{1}
\and
Loredana Spezzi\inst{2}
\and
Bruno Mer\'{\i}n\inst{3}
\and
Mar\'{\i}a Morales-Calder\'on\inst{1,4}
\and
Herv\'e Bouy\inst{3}
\and
David Barrado\inst{1}
\and
Jochen Eisl\"offel\inst{5}
}

   \institute{Laboratorio de Astrof\'{\i}sica Estelar y Exoplanetas, Dpto. Astrof\'{\i}sica, Centro de Astrobiolog\'{\i}a (LAEX-CAB), CSIC-INTA, ESAC Campus, P.O. Box 78, E-28261 Villanueva de la Ca\~nada, Madrid, Spain\\
   \email{belen@cab.inta-csic.es}
   \and
   Research and Scientific Support Department, European Space Agency (ESTEC), P.O. Box 299, 2200 AG Noordwijk, the Netherlands
   \and
   Herschel Science Centre, European Space Agency (ESAC), P.O. Box 78, E-28691 Villanueva de la Ca\~nada, Madrid, Spain
   \and
   Spitzer Science Center, California Institute of Technology, Pasadena, CA\,91125, USA
          \and
    Th\"uringer Landessternwarte, Sternwarte 5, D-07778 Tautenburg, Germany 
         }

   \date{Received; accepted}

 \abstract{
We combine published optical and near-infrared photometry to identify new low-mass candidate members in an area of about 0.64~deg$^2$ in Corona Australis with the $S$-parameter method.  Five new candidate members of the region are selected. They have estimated ages between 3 and 15~Myr and masses between 0.05 and 0.15~$M_{\odot}$. With {\em Spitzer} photometry we confirm that these objects are not surrounded by optically thick disks. However, one of them is found to display excess at 24~$\mu$m, thus suggesting it harbors a disk with an inner hole. With an estimated mass of 0.07$M_{\odot}$ according to the SED fitting, this is one of the lowest-mass objects reported to possess a transitional disk.  

Including these new members, the fraction of disks is about 50\% among the total Corona Australis population selected by the same criteria, lower than the 70\% fraction reported earlier for this region. Even so, we find a ratio of transitional to primordial disks (45\%) very similar to the value derived by other authors. This ratio is higher than for solar-type stars (5-10\%), suggesting that disk evolution is faster in the latter, and/or that the ``transitional disk'' stage is not such a short-lived step for very low-mass objects. However, this impression needs to be confirmed with better statistics.
 }{}{}{}{} 



   \maketitle

\section{Introduction}\label{sec:intro}

In the last two decades, our understanding of the formation and early evolution of low-mass stars and substellar objects has been remarkably improved, but there are still many unknowns regarding the exact mechanisms at play in these processes (see e.g. Whitworth et al. \cite{whitworth07} for a review). The census of the low-mass population for an increasing number of star-forming regions is important to address the problem of the universality of the initial mass function (IMF). The comparison of the disk fractions (that is, the number of sources with disks with respect to the total number of members) in clusters of similar and different ages also helps to constrain the timescales of disk evolution (e.g. Haisch et al. \cite{haisch01}). Moreover, the disk fraction and the spatial distribution of stars with and without disks provide important clues to understand the star-forming history and the disk evolution of a particular region. 

Most of our current knowledge about these issues comes from the study of star-forming regions and young clusters at distances within 100-300~pc from the Sun (see the reviews collected in Reipurth \cite{reipurth08}). However, in the vast majority of these studies the identification of cluster members is based on the detection of features related to accretion and circum(sub)stellar disks (e.g. H$\alpha$ emission, infrared excess), or indicative of strong activity (e.g. hard X-ray emission). Thus, these surveys are likely to be missing a fraction of the young low-mass population, especially objects that have dissipated all or most of their disks, that are in quiescence at the moment of the observations, 
and/ or whose activity level lies below the detection threshold of current X-ray surveys.

In a recent paper Comer\'on, Spezzi \& L\'opez Mart\'{\i} (2009, hereafter \cite{comeron09}) reported on a large-scale optical survey of the \object{Lupus} clouds,  selected to include most of the areas surveyed
in the {\em Spitzer} Legacy program {\em ``From molecular cores to planet-forming disks''} ({\em c2d}).
Combining their data with 2MASS near-infrared photometry, these authors developed an analysis procedure based on a dimensionless parameter $S$ that allowed them to easily identify very low-mass members of the dark clouds as an excess in the expected S-parameter distribution of field cool dwarfs and giants (see Sect.~\ref{sec:analysis} for details). In this way they were able to unveil a large population of objects belonging to Lupus I and III, which seems to be composed of very low-mass stars and brown dwarfs that have lost their inner disks on a timescale of a few Myr. The discovery of a substantial and even dominant population of thus far unnoticed members of one of our nearest star-forming regions stresses the important unknowns that still subsist in the observational characterization of young very low-mass objects. It also poses the important question of whether similar populations of yet unknown very low-mass members exist in other regions.

In this paper, we apply the $S$-parameter analysis to a set of optical and near-infrared observations of \object{Corona Australis}, another of the nearest regions with ongoing or recent intermediate- and low-mass star formation. Our aim is the identification of new very low-mass candidate members of this dark cloud. We focus on the core in the direction of the intermediate-mass star \object{R~CrA}, which contains a compact embedded stellar cluster known as the ``\object{R~CrA cluster}'' or ``the {\em\object{Coronet}} cluster'' (Taylor \& Storey \cite{taylor84}). With an age of about 3~Myr, its distance has been estimated to be within 50-170~pc  (see Neuh\"auser \& Forbrich \cite{neuhauser08} for a review). The young stellar population in this region has been studied with different techniques, including H$\alpha$ surveys (e.g. Marraco \& Ryndgren \cite{marraco81}), optical spectroscopy (e.g. Walter et al. \cite{walter97}), near-infrared mapping (e.g. Wilking et al. \cite{wilking92}), mid-infrared surveys (e.g. Olofsson et al. \cite{olofsson99}) and X-ray observations (e.g. Neuh\"auser et al. \cite{neuhauser00}; Forbrich \& Preibisch \cite{forbrich07}). Some brown dwarfs and brown dwarf candidates have been reported (Wilking et al. \cite{wilking97}; Fern\'andez \& Comer\'on \cite{fernandez01}; Bouy et al. \cite{bouy04}; L\'opez Mart\'{\i} et al. \cite{lm05}). Very recently, this cloud has also been the target of {\em Spitzer} observations (Sicilia-Aguilar et al. \cite{sicilia08}; L\'opez Mart\'{\i} et al. \cite{lm09}). 

The structure of the paper is as follows: In Section~\ref{sec:data} we present the data used in our analysis. Section~\ref{sec:analysis} summarizes the basis of the $S$-parameter formalism, discusses the contamination expected in our sample and describes the object selection. Section~\ref{sec:disc} is a discussion of the membership status and the properties of our new candidate members. Finally, in Section~\ref{sec:concl} we draw our conclusions.

\section{Optical and near-infrared data}\label{sec:data}

\begin{figure}[t]
   \centering
  \includegraphics[width=9cm]{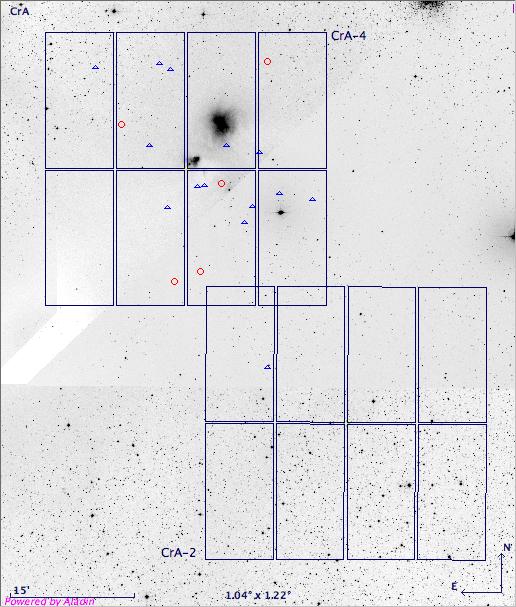}\hfill
      \caption{\footnotesize
	        Spatial location of the previously known candidate members (triangles) and the new candidates from this work (circles) within the Corona Australis dark cloud. Note that only those previous candidate members selected with the same criteria as our new candidates are plotted. The two WFI fields are also indicated. The objects are clustered around the intermediate-mass stars, which suggests that they belong to the {\em Coronet} cluster. 
	       }
         \label{fig:dist}
   \end{figure}

The analysis presented here is based on the same set of optical observations of the Corona Australis star-forming region reported in L\'opez Mart\'{\i} et al. (\cite{lm05}). The survey was performed with the WFI mosaic camera at the MPG/ESO 2.2m telescope of La Silla Observatory. It covered an area of about 0.64~deg$^2$, distributed in two fields of 34$^{\prime}\times$34$^{\prime}$, which are placed in the core of the {\em Coronet} cluster and on a region of lower cloud density to the South of it (Fig.~\ref{fig:dist}). For the present work, only the broadband observations in the $R$ and $I$ filters were considered, which cover the dynamical ranges $11<R<22$ and $12<I<23$. The survey is complete down to $R\simeq20$~mag and $I\simeq19$~mag. For details on the observations, data reduction and calibration, see L\'opez Mart\'{\i} et al. (\cite{lm05}) and references therein.

The optical data were cross-matched with $JHKs$ photometry from the 2MASS archive (Skrutskie at al. \cite{skrutskie06}), which is complete down to $J\simeq16.5$. For the present analysis, we require that the objects are detected in all five $RIJHKs$ bands. The total number of sources fulfilling this criterion amounts to 6136. Our sample is complete down to a mass of about $0.015M_{\odot}$ at a distance of 130~pc and an age of 3~Myr, according to the Lyon models (Baraffe et al. \cite{baraffe98}; Chabrier et al. \cite{chabrier00}).

\section{Analysis}
\label{sec:analysis}

\subsection{SED fitting}\label{sec:seds}

The spectral energy distributions (SEDs) of the objects in our optical-NIR catalog were fitted to the model spectra of Hauschildt et al. (\cite{hauschildt99}) and Allard et al. (\cite{allard00}) as explained in \cite{comeron09} (see also Spezzi et al. \cite{spezzi07}). The observed magnitude at each band is expressed as

\begin{equation}\label{eq:magfit}
m^{\lambda}_{obs}=M^{\lambda}_{synt}+A_V\cdot\frac{A_{\lambda}}{A_V}+S,
\end{equation}

\noindent
where $M^{\lambda}_{synt}$ is a synthetic magnitude in the same passband, computed from the model spectra, for a 1$R_{\odot}$ star at the same temperature; and S is a wavelength-independent scaling factor defined as

\begin{equation}\label{eq:spar}
S=5\log D(pc)-5\log R(R_{\odot})-5.
\end{equation}

\noindent
The ratio $A_V/A_{\lambda}$  is  a constant given by the adopted extinction law (Cardelli et al. \cite{cardelli89}, with $R_V=3.1$). A value of $\log g=4.0$, thought of being representative for young low-mass objects, was assumed for all the sources.

Assuming a certain $T_{eff}$ for each star, an equation of the form (\ref{eq:magfit}) was set for each passband and solved by least-squares to derive $A_V$ and $S$. The temperature that minimized the residual of the fit was then taken as the temperature of the object. The radius $R$ was computed from the corresponding value of $S$ for each source using Eq. (\ref{eq:spar}) and a distance value of $130\pm20$~pc to the Corona Australis star-forming region (Marraco \& Ryndgren \cite{marraco81}; Neuh\"auser \& Forbrich \cite{neuhauser08}). The luminosity $L$ was then computed with the relation $L=4\pi\sigma R^2T_{eff}^4$, where $\sigma$ is the Stefan-Boltzmann constant. Finally, the best-fitting $T_{eff}$ and the derived luminosity were used to estimate stellar masses and ages from the evolutionary models and tracks by Baraffe et al. (\cite{baraffe98}) and Chabrier et al. (\cite{chabrier00}), 
again assuming a distance of $130\pm20$~pc. 

We note that the use of a different distance would modify the estimated radii, luminosities and ages, but not the derived effective temperatures, extinctions and $S$-parameter values. For instance, a distance of 170~pc, as suggested by Knude \& H{\o}g (\cite{knude98}), would imply higher luminosities (by 0.23 in logarithmic scale), larger radii (by less than 0.5 for $R\leq1.5R_{\odot}$) and younger ages by about 2.5~Myr. However, because the ages derived using only photometry and isochrones, as in our case, intrinsically contain an uncertainty of a few Myr (e.g.   Baraffe et al. \cite{baraffe02}, \cite{baraffe09}), the distance uncertainty is not a major issue for our parameter estimates.

The residuals for the best fit give in turn an estimate of the quality of the solution. We found that 94\% of the objects have residuals below 0.2~mag, which is considered acceptable given the accuracy of the photometry. The robustness of the solution is supported by the distribution of $\chi^2$ values, which shows a clear minimum at the best-fitting value for each parameter, and a steep rise after the three best fits, especially for $T_{eff}$. The differences in the residuals  between the three best fits are $\leq 0.01$ in most cases.

As discussed by \cite{comeron09}, poor fits will be obtained for unresolved pairs and for variable stars 
(because the WFI and 2MASS observations are not simultaneous). In addition, the spectral energy distributions of some objects may be affected by the signatures of accretion and warm circumstellar dust in the visible (veiling and strong emission lines) and near-infrared (infrared excess), leading to an erroneous estimate of the temperature. In both situations, our fits tend to overestimate both the temperature and the extinction (see \cite{comeron09} for details).

The mean extinction error was $\Delta A_V=$0.15~mag according to the outcome of the fitting procedure, suggesting that most objects in our sample are located in the foreground rather than in the background of the dark cloud. Sicilia-Aguilar et al. (\cite{sicilia08}) provide independent extinction measurements for some cloud members, which are derived from their spectral types and near-infrared colors.  For the sources in common with that work, our values of $A_V$ tend to be higher than those provided by Sicilia-Aguilar et al. However, the difference is not larger than 0.5~mag for most of the objects and can be attributed to variability and/or to the use of a different extinction law.

   \begin{figure}[t]
   \centering
  \includegraphics[width=9cm]{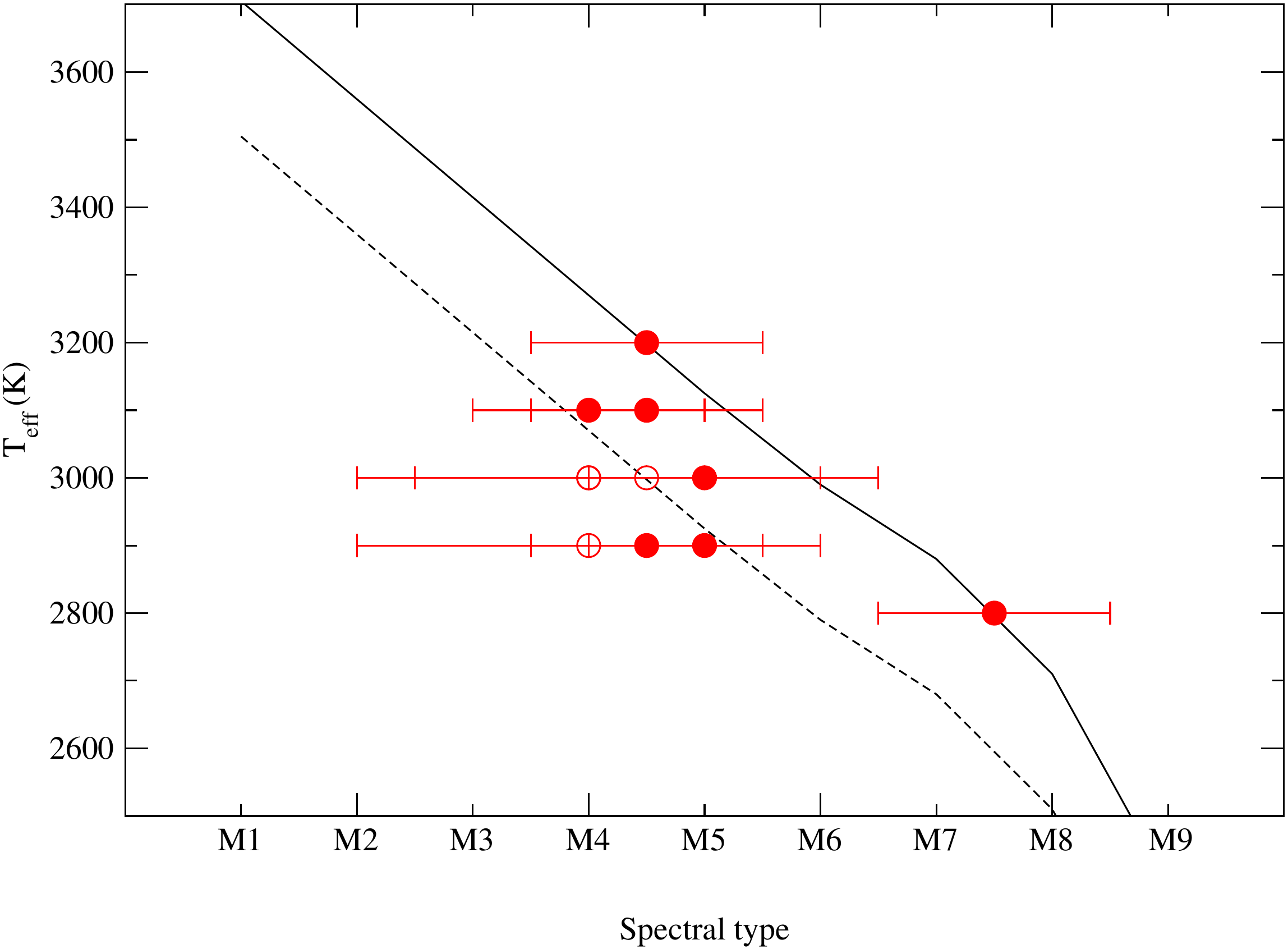}\hfill
      \caption{\footnotesize
	        $T_{eff}$ vs. spectral type for the previously known members and candidate members of the Corona Australis star-forming region listed in Table~\ref{tab:mem}.  Filled and open circles are used for spectral types derived from spectra (Sicilia-Aguilar et al. \cite{sicilia08}) and from narrow-band photometry (L\'opez Mart\'{\i} et al. \cite{lm05}), respectively. The solid line is the temperature scale for PMS objects from Luhman (\cite{luhman99}) and Luhman et al. (\cite{luhman05}). The dashed line is this scale shifted 200~K towards lower $T_{eff}$.
	       }
         \label{fig:Teff}
   \end{figure}

The uncertainty in the temperature determination is estimated from the comparison of the three best-fitting results for each object, which generally span a range of 200~K in effective temperature. We thus take 200~K as the error in our temperature values. This agrees with the results of Spezzi et al. (\cite{spezzi07}) and \cite{comeron09}, who estimated an error of 250~K on average from the comparison of their fitting results (obtained with the same procedure) with the effective temperatures derived from spectra for a sample of well-known members of Chamaeleon~II and Lupus, respectively.

In Fig.~\ref{fig:Teff}  we compare our $T_{eff}$ values with the published spectral types, derived from narrow-band photometry (L\'opez Mart\'{\i} et al. \cite{lm05}) or spectroscopy (Sicilia-Aguilar et al. \cite{sicilia08}) for a sample of previously known members and candidate members of Corona Australis, selected in Sect.~\ref{sec:sel} below and listed in Table~\ref{tab:mem}. We also plotted the temperature scale from Luhman (\cite{luhman99}) and its extension to late M-type objects by Luhman et al. (\cite{luhman05}). As seen in this figure, most of the objects are systematically cooler according to our fitting than the predictions of the Luhman scale. Indeed, other authors have suggested that the true temperatures of PMS stars are lower than the predictions of this temperature scale (e.g. Mohanty \& Basri \cite{mohanty03}; Barrado y Navascu\'es et al. \cite{byn04}), the differences amounting to 150-200~K. Indeed, we find that shifting the Luhman scale 200~K towards lower temperatures (dashed line in Fig.~\ref{fig:Teff}) provides a better agreement with our derived $T_{eff}$ values. This difference is still within our estimated error range.

\begin{table*}\label{tab:mem}
\caption{SED fitting results for the previously known candidate members of Corona Australis selected with our analysis}
\centering
\begin{tabular}{ l c c c c c c c}
\hline
\hline
Name & $A_V$ (mag) & $T_{eff}$ (K) & $S$ & $L (L_{\odot})$ & $R (R_{\odot})$ & $M (M_{\odot})$ & age (Myr) \\
\hline
  \object{CrA~205  }  & 0.04$\pm$0.14 & 3000 & 7.69$\pm$0.07 & 0.0103$\pm$0.0006 & 0.376$\pm$0.012 & 0.072 & 8.0\\
  \object{CrA~452  }  & 1.14$\pm$0.11 & 3400 & 6.07$\pm$0.05 & 0.0755$\pm$0.004  & 0.79$\pm$0.02   & 0.4   & 12.7\\
  \object{CrA~453  }  & 2.00$\pm$0.12 & 3200 & 7.39$\pm$0.06 & 0.0177$\pm$0.001  & 0.433$\pm$0.012 & 0.11  & 8.0\\
  \object{CrA~465  }  & 1.42$\pm$0.26 & 2800 & 7.89$\pm$0.13 & 0.0065$\pm$0.0008 & 0.34$\pm$0.02   & 0.03  & 1.1\\
  \object{CrA~468  }  & 0.12$\pm$0.15 & 3100 & 6.98$\pm$0.07 & 0.0226$\pm$0.0015 & 0.52$\pm$0.02   & 0.13  & 8.0\\
  \object{CrA~4107 }  & 0.73$\pm$0.15 & 3000 & 6.63$\pm$0.08 & 0.028$\pm$0.002   & 0.62$\pm$0.02   & 0.072 & 2.8\\
  \object{CrA~4108 }  & 0.16$\pm$0.16 & 2900 & 6.80$\pm$0.08 & 0.0205$\pm$0.0015 & 0.57$\pm$0.02   & 0.055 & 2.2\\
  \object{CrA~4109 }  & 0.45$\pm$0.13 & 3000 & 6.24$\pm$0.06 & 0.039$\pm$0.002   & 0.74$\pm$0.02   & 0.075 & 1.4\\
  \object{CrA~4110 }  & 0.94$\pm$0.16 & 3000 & 7.09$\pm$0.08 & 0.0170$\pm$0.0013 & 0.50$\pm$0.02   & 0.07  & 4.0\\
  \object{CrA~4111 }  & 0.22$\pm$0.22 & 2900 & 7.50$\pm$0.11 & 0.0107$\pm$0.0011 & 0.41$\pm$0.02   & 0.055 & 5.0\\
  \object{G-14     }  & 2.18$\pm$0.13 & 3100 & 7.30$\pm$0.07 & 0.0169$\pm$0.0010 & 0.452$\pm$0.014 & 0.09  & 6.4\\
  \object{G-102    }  & 1.07$\pm$0.17 & 2900 & 6.33$\pm$0.08 & 0.031$\pm$0.002   & 0.70$\pm$0.03   & 0.07  & 2.0\\
  \object{CrAPMS~3B}  & 0.33$\pm$0.15 & 2700 & 6.14$\pm$0.07 & 0.028$\pm$0.002   & 0.77$\pm$0.03   & 0.04  & 1.1\\
  \object{IRAC-CrA~3} & 1.73$\pm$0.25 & 2700 & 8.48$\pm$0.13 & 0.0033$\pm$0.0004 & 0.261$\pm$0.015 & 0.03  & 9.0\\
\hline
\end{tabular}
\tablebib{
L\'opez Mart\'{\i} et al. (\cite{lm05}, and \cite{lm09}); Sicilia-Aguilar et al. (\cite{sicilia08}); Neuh\"auser \& Forbrich (\cite{neuhauser08}); Meyer \& Wilking (\cite{meyer09})
}
\end{table*}

\subsection{Contamination}\label{sec:cont}

   \begin{figure}[t]
   \centering
  \includegraphics[width=9cm]{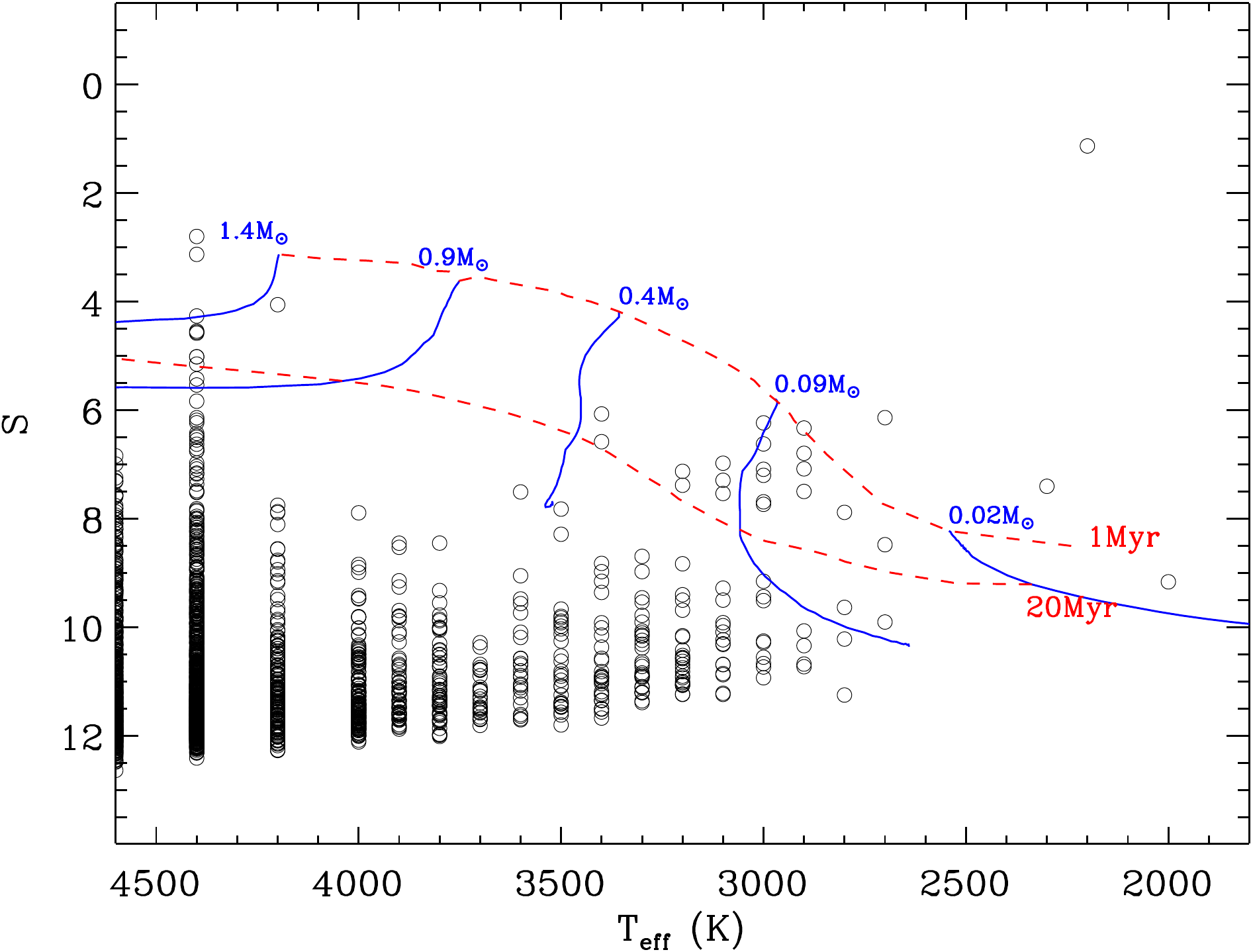}\hfill
      \caption{\footnotesize
	        ($T_{eff}$, $S$) diagram for those sources in our survey with a derived effective temperature below 4500~K. Mass tracks and isochrones from the models of Baraffe et al. (\cite{baraffe98}) are also plotted.	       }
         \label{fig:TeffS}
   \end{figure}

In Fig.~\ref{fig:TeffS} we show the ($T_{eff}$, $S$) diagram for the sources in our survey. We also plotted two isochrones from the models of Baraffe et al. (\cite{baraffe98}), corresponding to the expected location of objects of 1 and 20~Myr, and several mass tracks from the same models for masses between 1.4 and 0.02$M_{\odot}$. This diagram suggests that no Corona Australis members are present in our survey with temperatures between 4100 and 3400~K, as this range of $T_{eff}$ is devoid of objects in the area between and above the model isochrones. Indeed, cloud members with $T_{eff}> 3400$~K (spectral types earlier than about M3) are generally expected to be saturated in our optical survey ($I\lesssim 12$ for a cluster age of 3~Myr and a distance of 130~pc, according to the Baraffe et al. (\cite{baraffe98}) tracks).\footnote{ \footnotesize
As a matter of fact, a couple of known, highly extincted cluster members in this temperature range are actually present in our survey (namely CrA~466 and G-87, both with $A_V\gtrsim 8$~mag). However, our fits to their SEDs are not reliable, as they yield too high effective temperatures (around 4400~K), probably due to binarity or variability.}

   \begin{figure*}[t]
   \centering
  \includegraphics[width=16cm]{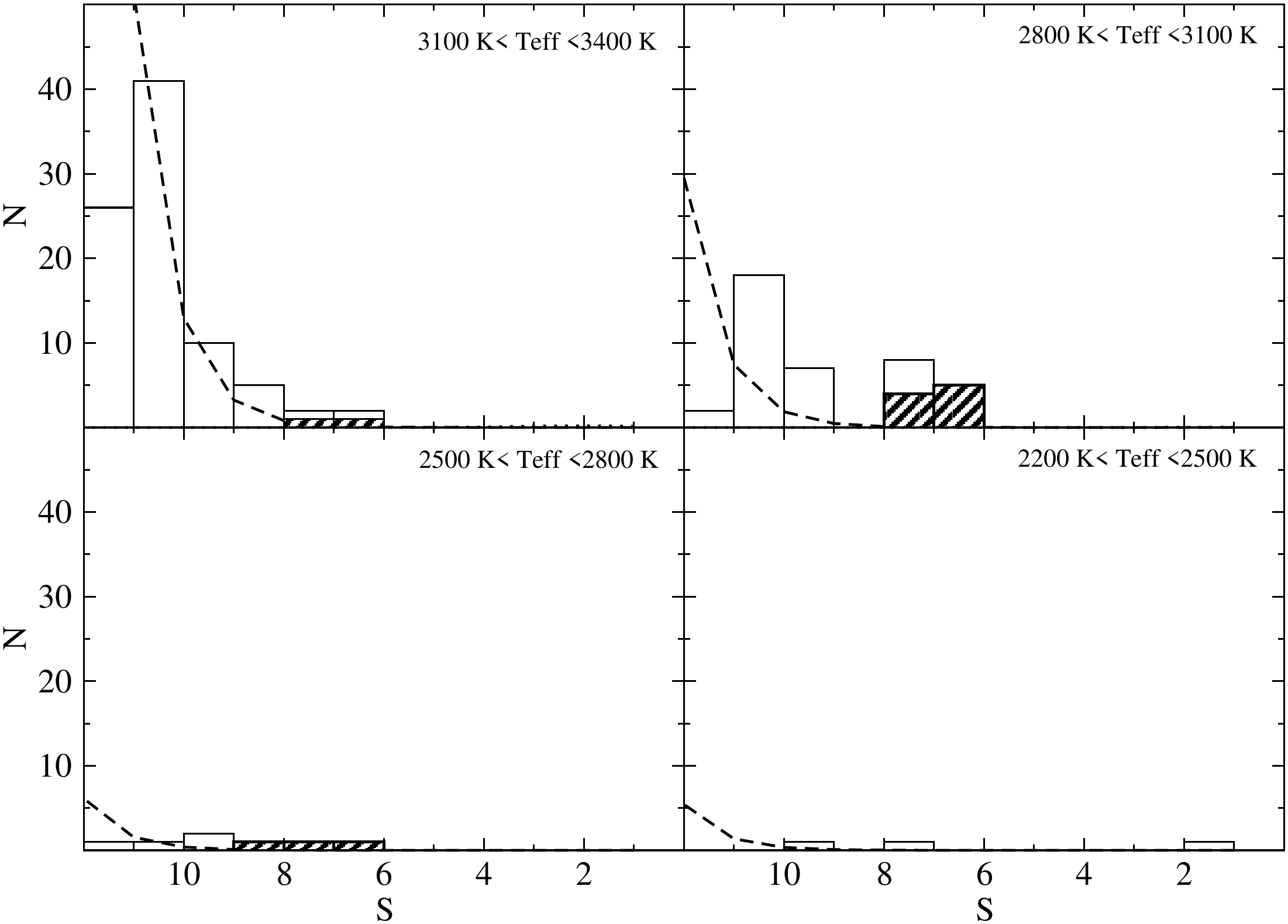}\hfill
      \caption{\footnotesize
	        Histograms of $S$ values at different temperature ranges in the Corona Australis star-forming region. The dashed lines indicate the expected contamination by field main sequence stars, and the hardly visible dotted lines at the bottom of each panel, the contamination by background cool giants. The presence of members of the star forming region is evident in those ranges of $S$ where contamination is expected to be negligible. The hashed histograms show the contribution of objects from previous studies to these excesses.	      
	         }
         \label{fig:hist}
   \end{figure*}

\begin{table*}\label{tab:cand_phot}
\caption{Optical and near-infrared photometry of our new candidate members}
\centering
\begin{tabular}{ l c c c c c }
\hline
\hline
Name &  $R$ &  $I$ &  $J$ & $H$ &  $Ks$ \\
\hline
\object{CrA~J190111.6-364532.0} & 18.067$\pm$0.004 & 15.779$\pm$0.005 & 13.375$\pm$0.026 & 12.496$\pm$0.028 & 12.129$\pm$0.026 \\
\object{CrA~J190139.1-370016.8} & 21.07$\pm$0.06   & 18.023$\pm$0.006 & 14.257$\pm$0.029 & 13.045$\pm$0.033 & 12.296$\pm$0.026 \\
\object{CrA~J190151.7-371048.4} & 16.451$\pm$0.002 & 14.500$\pm$0.003 & 12.326$\pm$0.030  & 11.423$\pm$0.033 & 11.047$\pm$0.030  \\
\object{CrA~J190207.6-371156.5} & 19.615$\pm$0.013 & 17.022$\pm$0.003 & 14.066$\pm$0.029 & 13.093$\pm$0.029 & 12.541$\pm$0.023 \\
\object{CrA~J190239.2-365311.1} & 17.928$\pm$0.005 & 15.654$\pm$0.005 & 13.190$\pm$0.024 & 12.471$\pm$0.022 & 12.060$\pm$0.023 \\
\hline\end{tabular}
\end{table*}

\begin{table*}\label{tab:cand}
\caption{SED fitting results for our new candidate members of Corona Australis}
\centering
\begin{tabular}{ l c c c c c c c}
\hline
\hline
Name &  $A_V$ (mag) &  $T_{eff}$ (K) &  $S$ &  $L (L_{\odot})$ & $R (R_{\odot})$ &  $M (M_{\odot})$ &  age (Myr) \\
\hline
CrA~J190111.6-364532.0 & 1.78$\pm$0.13 & 3000 & 7.20$\pm$0.06 & 0.0161$\pm$0.0009 & 0.471$\pm$0.013 & 0.07 & 4.5\\
CrA~J190139.1-370016.8 & 6.20$\pm$0.14 & 3200 & 7.13$\pm$0.07 & 0.0224$\pm$0.0014 & 0.487$\pm$0.016 & 0.13 & 8.0\\
CrA~J190151.7-371048.4 & 1.96$\pm$0.09 & 3400 & 6.58$\pm$0.05 & 0.047$\pm$0.002      & 0.627$\pm$0.013 & 0.3  & 14.3\\
CrA~J190207.6-371156.5 & 3.62$\pm$0.13 & 3100 & 7.54$\pm$0.06 & 0.0135$\pm$0.0007 & 0.404$\pm$0.011 & 0.1  & 10.0\\
CrA~J190239.2-365311.1 &  1.3 $\pm$0.2    & 2900 & 7.08$\pm$0.09 & 0.0157$\pm$0.0013 & 0.50$\pm$0.02   & 0.05 & 3.2\\
\hline\end{tabular}
\end{table*}

Furthermore, the low density of objects in the young locus together with the clear separation from the rest of sources in our survey suggest that contamination from field stars must be very low in this area of the diagram. Most of the remaining objects are placed well below the 20~Myr isochrone, indicating that they are foreground dwarfs characterized by high values of $S$. Background contamination would be caused by cool giants in the same range of effective temperatures as our objects of interest, but this contamination is expected to be very low due to the high galactic latitude of Corona Australis ($b\sim18^{\circ}$), and to the dark cloud itself. 

Following \cite{comeron09}, the foreground contamination was estimated by computing  the amount of main-sequence stars expected in our surveyed area for different temperature bins according to the local initial mass function (IMF) by Chabrier (\cite{chabrier03}). For $T_{eff}<3400$~K, the number of these objects expected to be found in the young locus in Fig.~\ref{fig:TeffS}  is three (3). The level of background contamination was estimated with the temperature and radii for cool giants provided by Fluks (\cite{fluks98}) and the volume density law from Wainscoat et al. (\cite{wainscoat92}). From this estimation only one (1) background contaminant is expected for $T_{eff}<3400$~K. Because we count 23 objects in the area of interest of  the ($T_{eff}$, $S$) diagram, the amount of expected contamination is about 17\%.

\subsection{Object selection}\label{sec:sel}

The procedure to identify candidate cloud members is the same as that outlined in \cite{comeron09} and is illustrated in Fig.~\ref{fig:hist}. For four intervals of effective temperature we plotted the expected S-parameter distribution of foreground and background stars. As seen in Fig.~\ref{fig:hist}, the number of foreground stars (dashed lines) decreases with decreasing values of $S$, and with decreasing $T_{eff}$.  On the other hand, the number of background stars (dotted lines) is negligible in all but the highest temperatures. 

In general, our histograms are close to the predicted numbers of contaminating stars, although incompleteness due to the limiting magnitudes affects the bins with the highest values of $S$. In those ranges of $S$ where the contamination is expected to be negligible, we interpret an excess of sources as objects that are probable members of the star-forming region. Based on the inspection of Fig.~\ref{fig:hist} and to minimize the number of contaminants, the following criteria were used to select candidate members of the Corona Australis star-forming region:

\begin{center}
\begin{tabular}{l c l}
$3100<T_{eff}<3400$~K, & & $S<8$; \\
$2800<T_{eff}<3100$~K, & & $S<9$; \\
$2500<T_{eff}<2800$~K, & & $S<10$;\\
$2200<T_{eff}<2500$~K, & & $S<10$.
\end{tabular}
\end{center}

In this way, 4, 13, 5 and 3 candidates respectively were selected in each temperature bin, amounting to 22 in total. The objects were then visually inspected to reject bad detections either due to bad pixels (two objects) or to source confusion in the cross-match (one object). Sources for which the fits derived ages older than expected for Corona Australis members were also rejected as likely contaminants;  concretely, we discarded two objects whose derived ages were older than 100~Myr. For another source, the SED fitting suggests an age of 18~Myr, which we consider to be slightly too old to be a member of Corona Australis. The residual in the best fit for this latter source is relatively high ($\chi^2\sim0.4$) due to its faintness (and hence large photometric errors) in the R-band ($R>20$~mag). Because we do not have any additional information to support membership for this object, we chose to be conservative and reject it as a cluster member as well.

This left us with 19 objects in the list of selected candidate members of Corona Australis. Thirteen (13) of them had been presented in previous works as candidate members of the region, and eight of these candidates had been confirmed as members through optical or near-infrared spectroscopy (see references in Table~\ref{tab:mem}). Another object (IRAC-CrA~3) had been selected by us as a candidate class II member of Corona Australis in a parallel study based on {\em Spitzer} photometry (L\'opez Mart\'{\i} et al. \cite{lm09}). Our fitting results for these fourteen objects are listed in Table~\ref{tab:mem}: They have effective temperatures between 2700 and 3400~K and masses betweeen 0.03 and 0.4$M_{\odot}$. Only two objects have masses above 0.1$M_{\odot}$. The optical extinction varies from 0 to 2.2~mag. 

Together with the previously known candidates, our list still contains five objects that are likely members of the Corona Australis region, all of them with ages younger than 15~Myr according to the SED fitting, and none with reported detections in previous works. They have effective temperatures and masses within the same range as the previously known objects, but tend to be somewhat more extincted, with $A_V$ values in the range 1-4~mag and one object with $A_V>6$~mag. The optical and near-infrared photometry for these new candidate members is given in Table~\ref{tab:cand_phot}, and their parameters from the SED fitting are listed in Table~\ref{tab:cand}. 

Note that (excluding the two objects rejected due to photometry problems and the one cross-matching misidentification) we classified three sources as likely contaminants, which agrees very well with our estimation from Sect.~\ref{sec:cont}. On the other hand it must be remarked that we are missing members of the region whose $S$-parameter values are in the ranges affected by contamination in the histograms of Fig.~\ref{fig:hist}. Indeed, some excess above the expected foreground contamination is observed in some bins to the left of our $S$ cutoff value for $T_{eff}<$3100~K (upper panels of Fig.  \ref{fig:hist}), where we would expect to find more cluster members. In addition, as explained in Sect.~\ref{sec:seds}, wrong fits would be obtained for objects affected by variability, binarity, strong accretion or infrared excess, and these sources would be wrongly placed in the ($T_{eff}$,$S$) diagram of Fig.~\ref{fig:TeffS}. As a consequence, cloud members with primordial disks may be slightly underrepresented in our selected sample.

In particular, other candidate members {\bf from} the studies of Fern\'andez \& Comer\'on (\cite{fernandez01}; 1 object), L\'opez Mart\'{\i} et al. (\cite{lm05}; 3 objects), Forbrich et al. (\cite{forbrich07}; 2 objects) and Sicilia-Aguilar et al. (\cite{sicilia08}; 2 objects) are detected in our survey with photometry in all $RIJHKs$ bands, but are not selected with our method. Hence, they are not listed in Table~\ref{tab:mem} nor in Fig.~\ref{fig:dist} (but they are included in Figs.~\ref{fig:TeffS} and \ref{fig:hist}). For the previously known candidate members present in our survey, our success rate in recovering them is thus about 65\%, or 72\% considering only objects whose membership had been confirmed by spectroscopy; these percentages agree well with the 30\% of members with poor fits found in the Lupus clouds (\cite{comeron09}). For the same reasons there could be some yet unknown members of CrA which are not selected by our criteria.

\section{Discussion}
\label{sec:disc}

\subsection{Membership of the candidates}\label{sec:status}

   \begin{figure}[t]
   \centering
  \includegraphics[width=9cm]{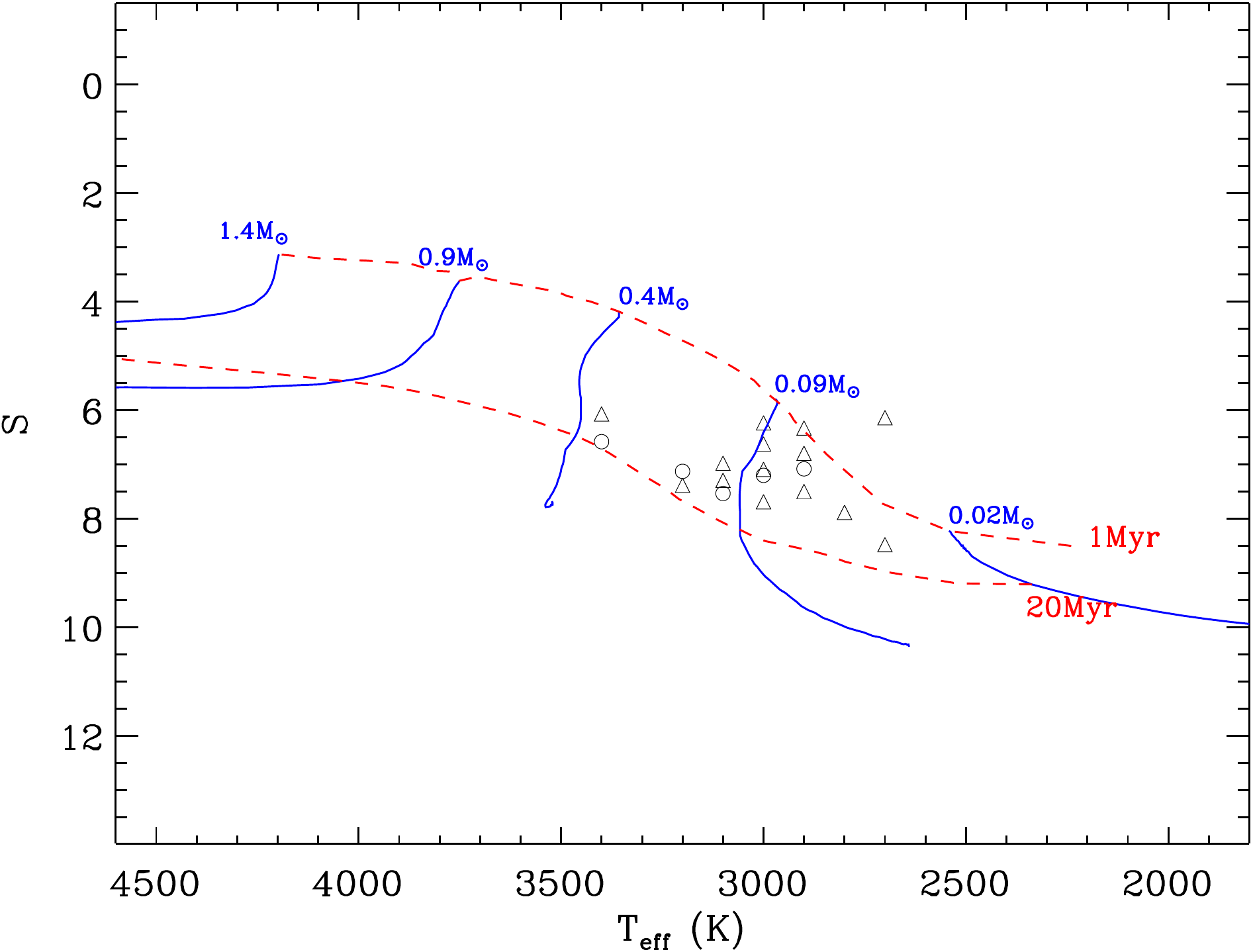}\hfill
      \caption{\footnotesize
	        Same as Fig.~\ref{fig:TeffS}  for the previous candidate members (triangles) and new candidate members from this work (circles) selected in Sect.~\ref{sec:sel}. 
	       }
         \label{fig:TeffSmem}
   \end{figure}

   \begin{figure}[ht]
   \centering
  \includegraphics[width=9cm]{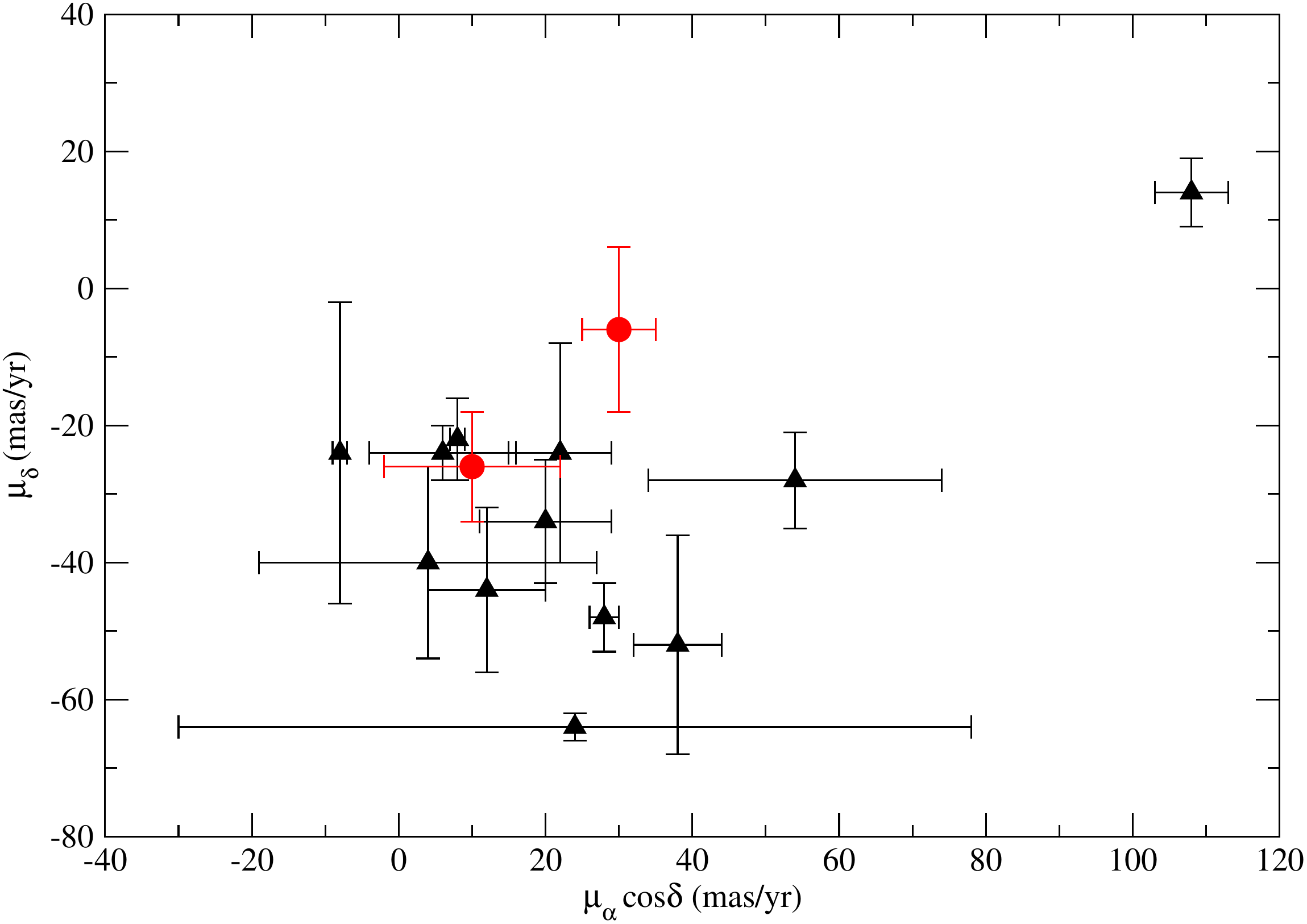}\hfill\\
  \includegraphics[width=9cm]{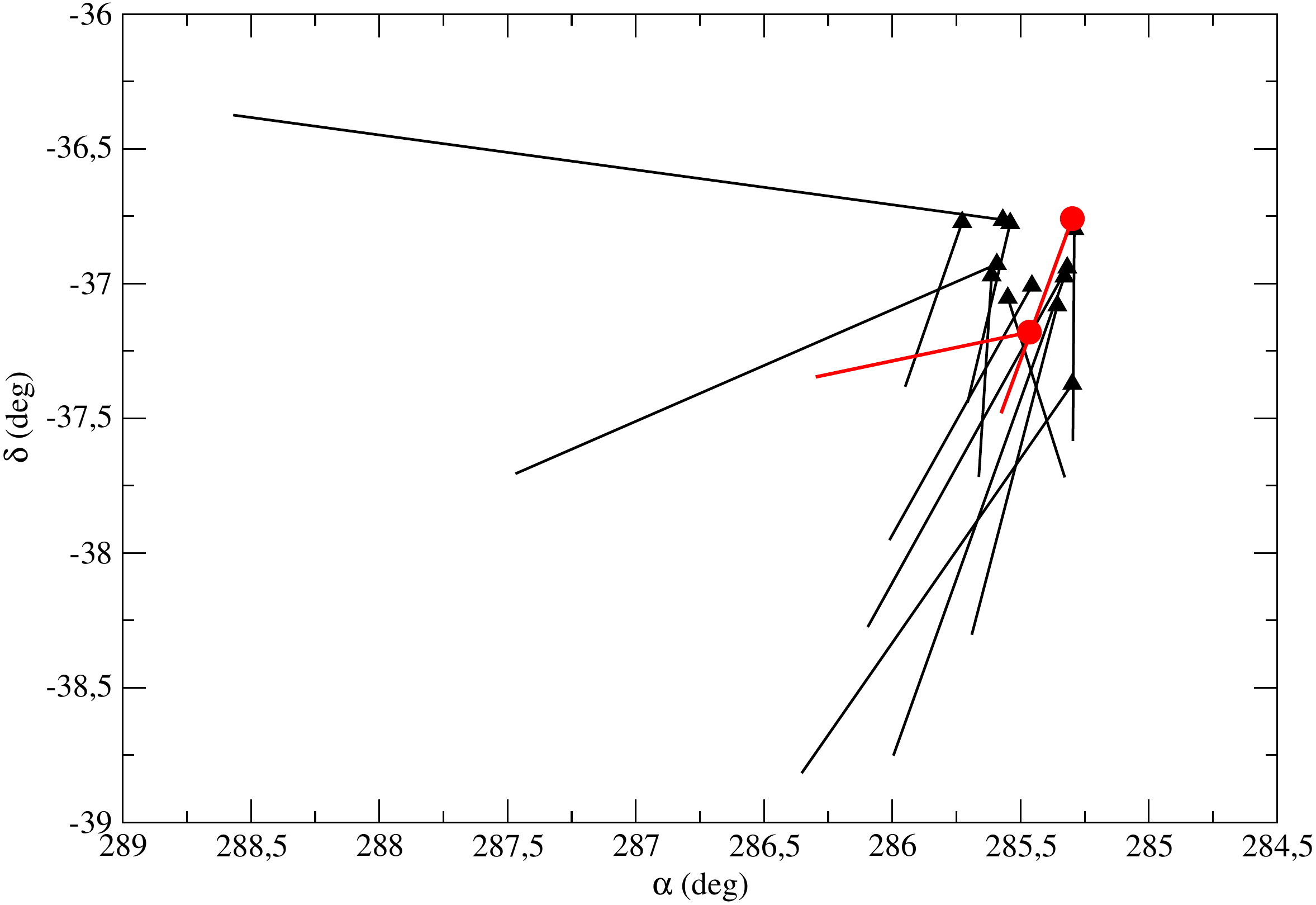}\hfill
      \caption{\footnotesize
	      {\em Upper panel:}  $\mu_{\delta}$ versus $\mu_{\alpha}\cos\delta$ from the USNO-B1 catalogue for {\em Coronet} candidate members from previous works (triangles) and our new candidate members (circles).  
	      {\em Lower panel:} Spatial location of the same sources. The lines indicate their expected displacement in 10$^5$~yr.
	       }
         \label{fig:pm}
   \end{figure}

\begin{table*}\label{tab:pminfo}   
\centering
\caption{Proper motion information for CrA members and candidate members}
\begin{tabular}{l c c c c c}
\hline
\hline
Name &  USNO-B1 & $\mu_{\alpha}\cos\delta$ (mas/yr) &  $\mu_{\delta}$ (mas/yr) & P &  Nobs \\
\hline
  CrA~205                           & 0526-0958780 &   38$\pm$6  & $-$52$\pm$16 & 9 & 4\\
  CrA~466                           & 0530-0881950 &   24$\pm$54 & $-$64$\pm$2  & 7 & 3\\
  CrA~468                           & 0529-0905310 &   20$\pm$9  & $-$34$\pm$9  & 9 & 4\\
  CrA~4107                          & 0532-0801755 & 	8$\pm$1  & $-$22$\pm$6  & 9 & 4\\
  CrA~4108                          & 0532-0800872 & 	6$\pm$10 & $-$24$\pm$4  & 9 & 4\\
  CrA~4109                          & 0532-0800993 &  108$\pm$5  &    14$\pm$5  & 9 & 4\\
  CrA~4110                          & 0530-0881945 &   28$\pm$2  & $-$48$\pm$5  & 9 & 4\\
  CrAPMS~3AB  		            & 0530-0882190 &   54$\pm$20 & $-$28$\pm$7  & 9 & 4\\
  $[$FP2007$]$ J190149.35-370028.6  & 0529-0905310 &   20$\pm$9  & $-$34$\pm$9  & 9 & 4\\
  G-1		                    & 0530-0882197 &   22$\pm$7  & $-$24$\pm$16 & 9 & 4\\
  G-14                              & 0529-0905404 & $-$8$\pm$1  & $-$24$\pm$22 & 7 & 4\\
  G-102                             & 0529-0905023 &   12$\pm$8  & $-$44$\pm$12 & 9 & 4\\
  RX J1901.1-3648		    & 0532-0800238 & 	4$\pm$23 & $-$40$\pm$14 & 9 & 4\\
  \hline
  CrA~J190151.7-371048.4            & 0528-0928314 &   30$\pm$5  &  $-$6$\pm$12 & 9 & 4\\
  CrA~J190111.6-364532.0            & 0532-0800263 &   10$\pm$12 & $-$26$\pm$8  & 9 & 4\\
\hline
\end{tabular}
\tablebib{
Neuh\"auser et al. (\cite{neuhauser00}); L\'opez Mart\'{\i} et al. (\cite{lm05}); Forbrich \& Preibisch (\cite{forbrich07}); Sicilia-Aguilar et al. (\cite{sicilia08}); Neuh\"auser \& Forbrich (\cite{neuhauser08}); this work
}

\end{table*}

We now discuss the membership status of our five objects. Figure~\ref{fig:TeffSmem} is a ($T_{eff}$, $S$) diagram analogous to Fig.~\ref{fig:TeffS}, but showing only the location of the members and candidate members of Corona Australis selected in this work. 

The spatial distribution of our sources is shown in Fig.~\ref{fig:dist}. Both the previously known candidate members from Table~\ref{tab:mem} and the new candidate members from Table~\ref{tab:cand} are located in the northern WFI field; the only exception is the candidate member CrA~205 from L\'opez Mart\'{\i} et al. (\cite{lm05}), which is placed in the southern field. This distribution supports the membership of our new candidates to the {\em Coronet} cluster around the intermediate-mass star R~CrA.

Two of our candidates,  CrA~J190111.6-364532.0 and CrA~J190151.7-371048.4, have a USNO-B counterpart within $2^{\prime\prime}$. We compare their proper motions as reported in this catalog with those of previously known members of the {\em Coronet} cluster (see Table~\ref{tab:pminfo}). We considered only USNO-B measurements with a high probability estimator ($\geq$0.7) and measurements in at least three epochs (Monet et al. \cite{monet03}). The proper motion data are summarized in Table~\ref{tab:pminfo}. 

With only 14 objects in total it is not possible to perform a statistically meaningful membership analysis based on proper motion. The upper panel of  Fig.~\ref{fig:pm} shows  $\mu_{\delta}$ versus $\mu_{\alpha}\cos\delta$ for all sources. In the lower panel we have plotted the current spatial locations of the sources and their expected displacement in 10$^5$~yr. While we cannot draw any firm conclusion on their membership, this analysis shows that at least one of the candidates, CrA~J190111.6-364532.0, has a proper motion consistent with other members. For CrA~J190151.7-371048.4 there is at least some marginal agreement (within the errors) in its proper motion components with respect to those of the majority of cluster members plotted in Fig.~\ref{fig:pm}. Inspection of the optical images shows that this object has two close visual companions, located about 3$^{\prime\prime}$ to the NE and 2$^{\prime\prime}$ to the NW respectively, a fact that may be affecting the proper motion measurements.

It is remarkable that one source, namely CrA~4109, displays clearly larger proper motion than the rest of probable cluster members plotted in Fig.~\ref{fig:pm}, with $\mu\simeq109$~mas/yr. The direction of its expected displacement is also significantly different. Given the low object density in our surveyed region, source confusion seems unlikely. Moreover, the USNO-B $R$ and $I$ photometry (16.07 and 13.98~mag, respectively) is fully consistent with the WFI photometry for this object ($R=15.99$ and $I=13.98$; L\'opez Mart\'{\i} et al. \cite{lm05}). Hence we are quite confident that our identification of CrA~4109 with this proper motion source is correct. On the other hand, although there is no definitive confirmation of its membership to Corona Australis, there are several indications in the literature (H$\alpha$ emission, mid-infrared photometry and spectroscopy) suggestive of CrA~4109 being a young object (L\'opez Mart\'{\i} et al. \cite{lm05}; Sicilia-Aguilar et al. \cite{sicilia08}).

In Fig.~\ref{fig:dist} CrA~4109 is located on the outskirts of the cloud core in an area of relatively low extinction. A possibility would thus be that this object was ejected from its birth site due to dynamical interactions with other cluster members, maybe within a multiple system.  
We note that two more sources (CrA~4107 and CrA~4108) are located in the same area as CrA~4109, but their proper motions agree very well with those of the majority of cluster members.

\subsection{Disks?}\label{sec:disks}

   \begin{figure*}[t]
   \centering
  \includegraphics[width=40cm, bb=130 200 900 600]{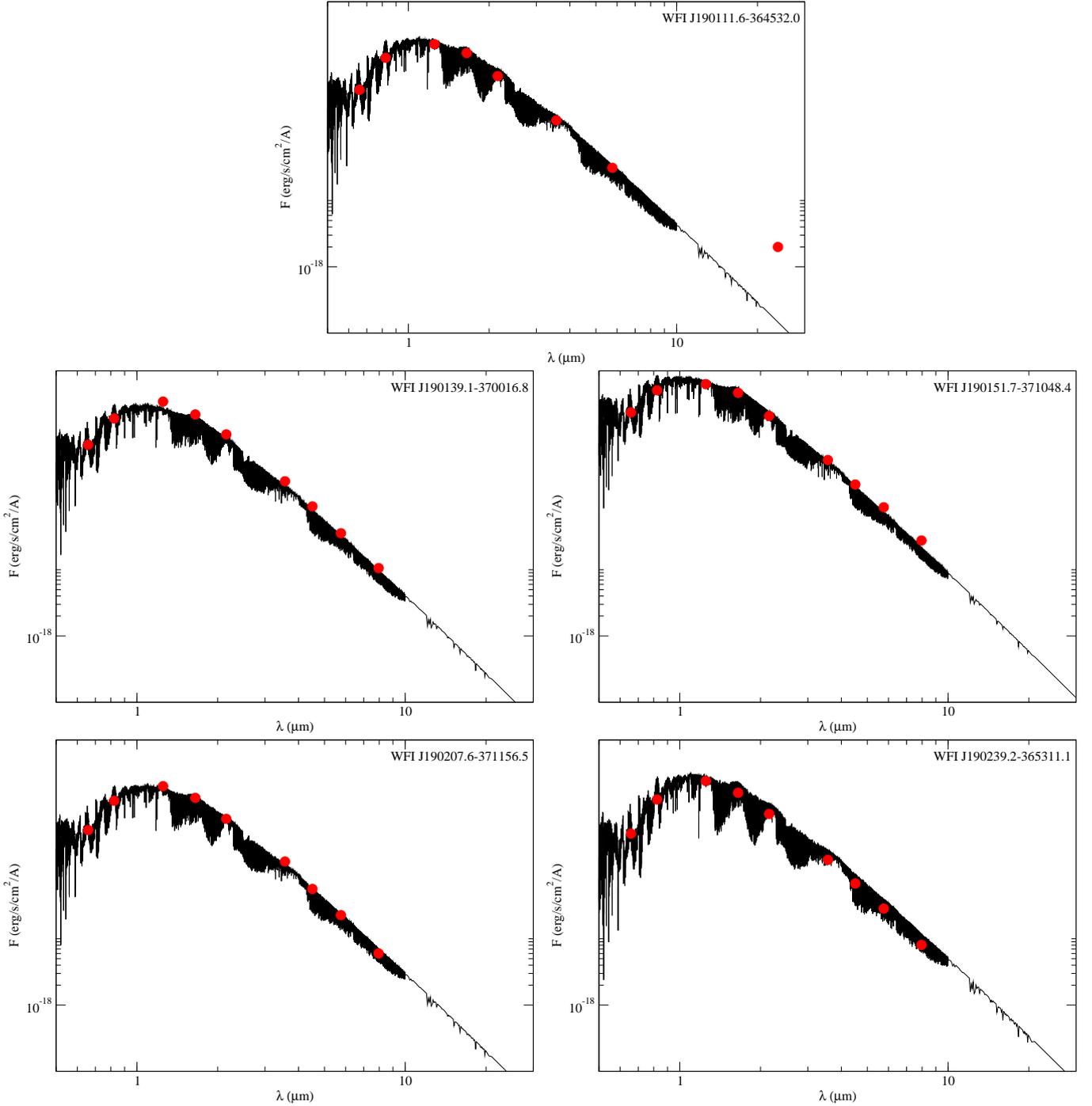}\hfill
      \caption{\footnotesize
	        Spectral energy distributions of our new candidate members of Corona Australis (red dots), from the optical to the mid-infrared, compared to their corresponding best-fitting photospheres according to the SED analysis (see model parameters in Table~\ref{tab:cand}).
	       }
         \label{fig:cand_seds}
   \end{figure*}

\begin{table*}\label{tab:cand_spitzer}
\caption{{\em Spitzer} photometry of our new candidate members}
\centering
\begin{tabular}{ l c c c c c c c c}
\hline
\hline
Name & [3.6] & [4.5] &  [5.8] & [8.0] &  [24.0] & $\alpha_{IRAC}$ &  Class &  Disk$^{a}$ \\
\hline
CrA~J190111.6-364532.0 & 11.653$\pm$0.004 &                                  & 11.375$\pm$0.015 &                                    & 8.23$\pm$0.075 &                   &     &   \\
CrA~J190139.1-370016.8 & 11.653$\pm$0.005 & 11.48$\pm$0.006 & 11.415$\pm$0.016 & 11.392$\pm$0.026 &            n.d.             & $-$2.556 & III & A/N \\
CrA~J190151.7-371048.4 & 10.553$\pm$0.003 & 10.44$\pm$0.003 & 10.265$\pm$0.007 & 10.182$\pm$0.008 &            n.d.             & $-$2.397 & III & A/N \\
CrA~J190207.6-371156.5 & 11.893$\pm$0.004 & 11.86$\pm$0.006 & 11.795$\pm$0.017 & 11.902$\pm$0.03   &            n.d.             & $-$2.836 & III & N \\
CrA~J190239.2-365311.1 & 11.673$\pm$0.004 & 11.54$\pm$0.005 & 11.455$\pm$0.015 & 11.492$\pm$0.018 &            n.d.             & $-$2.633 & III & N \\
\hline
\end{tabular}
\begin{flushleft}
{\bf Notes.}
n.d.$=$not detected; blank$=$not observed \\
$^{(a)}$ A$=$thin disk; N$=$no disk (see discussion in text)
\end{flushleft}
\end{table*}

Given that our candidates from Table~\ref{tab:cand} had not been detected in previous surveys (the vast majority of them looking for signatures of accretion or disks), they are expected to be diskless, or to have at most optically thin disks. To check this, we cross-matched our optical-NIR catalog with the {\em Spitzer} IRAC and MIPS 24~$\mu$m observations from the study reported in L\'opez Mart\'{\i} et al. (\cite{lm09}; see that work for details on the reduction and analysis of these data). 

The mid-infrared photometry for our new candidates is summarized in Table~\ref{tab:cand_spitzer}. Four of them have $[3.6]-[4.5]$ and $[5.8]-[8.0]$ colors close to zero, as expected for class III sources (e.g. Allen et al. \cite{allen04}; Hartmann et al. \cite{hartmann05}). With the criterion proposed by Lada  et al. (\cite{lada06}) to make a classification of the disks based on the value of the SED slope $\alpha_{IRAC}$ in the 3.6-8.0~$\mu$m range, we find that two sources, CrA~J190239.2-365311.1 and CrA~J190207.6-371156.5, are classified as diskless ($\alpha_{IRAC}<-2.56$), while CrA~J190151.7-371048.4 and CrA~J190139.1-370016.8  could be surrounded by anemic (or thin) disks ($-2.56<\alpha_{IRAC}<-1.80$). We remark though that the slope values of the last two objects ($-2.556$ and $-2.397$, respectively) are very close to the defined limiting value between anemic disks and no disks. Furthermore these limits are not absolute, because they depend slightly on spectral type and (more strongly) on photometric errors (Lada  et al. \cite{lada06}; Hern\'andez et al. \cite{hernandez08}). Given the dispersion on the slope of the field (thus, diskless) objects in our survey ($\sigma\sim0.1$), we consider CrA~J190151.7-371048.4 and CrA~J190139.1-370016.8 as diskless. The remaining source (CrA~J190111.6-364532.0) could not be classified according to these schemes, because it was not observed at 4.5 and 8.0~$\mu$m.

The SEDs of our candidates are shown in Fig.~\ref{fig:cand_seds} compared to the best-fitting model photosphere for each object. We note that the error bars are smaller than the symbol size in the plots due to the low flux uncertainties. From visual inspection, the four sources classified as diskless display nearly photospheric SEDs in the whole range in which they are detected (up to 8~$\mu$m).
 
The source CrA~J190111.6-364532.0 is the only one with a counterpart at 24~$\mu$m. As seen in Fig.~\ref{fig:cand_seds}, its SED has the characteristic shape of a star that has cleared its inner disk, showing a clear excess over the photosphere only after 10~$\mu$m ---a so-called ``transitional disk'' (e.g. Calvet \& D'Alessio \cite{calvet01}; Calvet et al. \cite{calvet05}).  This source would have a mass of about 0.07$M_{\odot}$ according to the SED fitting, which would make it one of the lowest-mass objects reported to possess a transitional disk to date. 

\subsection{Disk fraction and transition disks}\label{sec:dfrac}

Among the previously known objects selected in Sect.~\ref{sec:sel} and listed in Table~\ref{tab:mem} there are five class II sources (G-14, CrA~465, CrA~4107, CrA~4110 and IRAC-CrA~3), four class III sources (CrA~453, CrA~468, CrA~4108 and CrA~4111)\footnote{\footnotesize
CrA~4111 was classified as a transitional object (in the sense in which we are using the term here) by Sicilia-Aguilar et al. (\cite{sicilia08}), who report the detection of a 24~$\mu$m excess for this source. Our own analysis of the {\em Spitzer} data, however, does not confirm this result: We do not find any counterpart for CrA~4111 at 24~$\mu$m. Visual inspection of the MIPS 24~$\mu$m mosaic shows that this object is located in an area of nebulosity, which could have yielded a spurious detection.}
and three transitional objects (G-102, CrA~205 and CrA~4109) according to Sicilia-Aguilar et al. (\cite{sicilia08}) and our own analysis (L\'opez Mart\'{\i} et al. \cite{lm09}). Note that by ``transitional object'' we mean here a star or brown dwarf harboring a disk with an inner hole, as suggested by a 24~$\mu$m excess larger than at 8~$\mu$m or shorter wavelengths. No classification is available for CrAPMS~3B, because this is the low-mass companion of a PMS star and is unresolved in the IRAC images, nor for CrA~452, which lies outside the area covered by the IRAC observations.

If we now add our five new candidates to the sample, the final census amounts to five sources with optically thick disks, eight sources without disks, and four sources with transitional disks for a total of 17 classified objects selected following the same criteria. Taking both primordial and transitional disks together, the disk fraction in our sample amounts to 53$\pm$18\%, lower than the 70\% reported by Sicilia-Aguilar et al. (\cite{sicilia08}), due to the increase of diskless objects.

The {\em primordial} disk fraction in our sample is 30$\pm$13\%, lower than the estimations of Haisch et al. (\cite{haisch01}) and Mamajek (\cite{mamajek09}) for a 3~Myr cluster (the usually quoted age for the {\em Coronet} cluster), but not far from their given value of about 50$\pm$10\% when the errors are considered. We note, though, that a relatively large age spread is observed among the objects in our sample, according to the SED fitting results. The derived primordial disk fraction is only slightly higher, within the errors, than that obtained by Haisch et al. (\cite{haisch01}) for the 5~Myr \object{NGC~2362} cluster ($12\pm4\%$), which agrees quite well with the median age of our sample (5~Myr). However, this result should not be interpreted as a new estimation of the cluster age, for several reasons. The first one is of course the low number of objects in our sample and its limited mass range. Second, we caution that individual ages derived from theoretical SED fitting are affected by the many uncertainties still present in the models and may even depend on the accretion history of the objects (e.g. Baraffe et al. \cite{baraffe02}, \cite{baraffe09}; Mayne \& Naylor \cite{mayne08}). In addition, as explained in Sect.~\ref{sec:analysis}, the objects with strongest accretion (thus probably the youngest) are likely to be missed by our selection, which produces a slightly older sample than the mean age of the cluster. The standard deviation of our sample age (3.9) accounts well for these uncertainties.

On the other hand, the ratio of transitional to primordial disks (45$\pm$22\%) agrees quite well with the 50\% value reported by Sicilia-Aguilar et al. (but see also Ercolano et al. \cite{ercolano09}). As already noted by these authors, this ratio is higher than observed in other regions of similar age (around 5-10\%;  see e.g. Mer\'{\i}n et al., \cite{merin09}, for a discussion of the transitional disk ratios in the {\em c2d} clouds). However, the majority of these studies targeted higher mass stars than considered in Sicilia-Aguilar et al. (\cite{sicilia08}) and the present study. The larger number of transitional disks among low-mass stars and brown dwarfs with respect to solar-type stars could then be interpreted as a hint of a faster disk evolution in the former, as suggested by previous works (e.g. Sterzik et al. \cite{sterzik04}; Bouy et al. \cite{bouy07}). It also suggests that the ``transitional'' stage is not necessarily a short-lived step in the evolution towards a protoplanetary or debris disk, at least for very low-mass objects. However, our sample is too small to derive any statistically meaningful conclusion on this issue.  

We remark that the inclusion of the two unclassified objects (CrAPMS~3B and CrA~452) in this census will not alter the measured disk fraction by more than 5\% in either sense (higher or lower disk fraction), nor will it change the ratio of transitional to primordial disks by more than 10\%, and hence will not substantially affect these conclusions. On the other hand, as explained in Sect.~\ref{sec:seds}, cloud members with primordial disks may be slightly underrepresented in our selected sample. However, even in the most pessimistic case (that would be, if the whole 30\% of badly fitted sources estimated in Sect.~\ref{sec:seds} were surrounded with primordial disks), the disk fractions would still be within the uncertainties of our estimations.

\section{Conclusions}
\label{sec:concl} 

We performed an analysis of optical and near-infrared data of the Corona Australis star-forming region based on the $S$-parameter formalism. In our surveyed area of $\sim0.64$~deg$^2$, we identified fourteen previous candidate members and five new candidate members of this dark cloud.

The new candidates have estimated effective temperatures between 2900 and 3400~K, corresponding to masses between 0.05 and 0.13$M_{\odot}$. They are thus very low-mass stars and massive brown dwarfs.  Their ages span between 3 and 15~Myr, which is consistent with the reported age spread in Corona Australis ($\sim$1-10~Myr), given the uncertainties in the models, in the distance to the cloud and in the SED fitting procedure. The membership of these objects to the star forming region is further supported by their spatial distribution and, when available, proper motion information. The source for which membership is more uncertain is CrA~J190151.7-371048.4, the oldest star in our sample according to the SED fitting results. 

The SEDs of four of our new candidates are nearly photospheric. The exception is CrA~J190111.6-364532.0: This source displays excess at 24~$\mu$m, which is suggestive of a transition disk with an inner hole. With an estimated mass of 0.07$M_{\odot}$, this is one of the lowest-mass objects reported to possess such a disk. 

We calculated the disk fraction of the Corona Australis population selected with our method, which is 50\%. This value is lower than the one reported in a previous study by Sicilia-Aguilar et al. (\cite{sicilia08}). The ratio of transitional to primordial disks (45\%) though agrees well with the fraction of 50\% reported by Sicilia-Aguilar et al. (\cite{sicilia08})  and is remarkably higher than the value measured in other clusters of similar age. This suggests that transitional disks around brown dwarfs may have longer lifetimes than around low-mass stars. However, this impression should be confirmed with a larger, more statistically meaningful sample.

The results from this work stress the need to properly characterize the diskless population of a region to derive meaningful disk fractions. This is especially important to understand the dependence of the disk fractions with the mass of the central objects, in particular for the very low-mass population. Even in a relatively small survey like the one reported here and in a well-studied and relatively low-density region like Corona Australis, 25\% of the total number of cloud members selected with our method had not been detected in previous studies based on accretion, disk or activity signatures. An analogous study of an eventual larger-scale survey covering most of the dark cloud would probably identify a significant number of young very low-mass objects belonging to this star-forming region.

\begin{acknowledgements}

We are very grateful to F. Comer\'on for his help with the data analysis and his comments on an early draft of this paper. We also thank A. Bayo for useful discussions.

This work was partially funded by the Spanish MICINN through grants Consolider-CSD2006-00070 and ESP2007-65475-C02-02. It also had funding from the Madrid regional government through grant CAM/PRICIT-S2009ESP-1496.

This research is based on observations collected at the European Southern Observatory, on La Silla (Chile), and on data products from the Two Micron All Sky Survey, which is a joint project of the University of Massachusetts and the Infrared Processing and Analysis Center/California Institute of Technology, funded by the National Aeronautics and Space Administration and the National Science Foundation.  It also used observations made with the {\em Spitzer} Space Telescope, operated by the Jet Propulsion Laboratory, California Institute of Technology, under a contract with NASA. 

This publication greatly benefited of the use of the SIMBAD database and VIZIER catalogue service, both operated at CDS (Strasbourg, France). We used the VO-compliant tools Aladin, developed at CDS, TOPCAT, currently developed within the AstroGrid project, and VOSA, developed under the Spanish Virtual Observatory project, and supported by the Spanish MICINN through grant AyA2008-02156. 

\end{acknowledgements}


\end{document}